\documentclass[aps,prl, twocolumn,footinbib,superscriptaddress,floatfix]{revtex4-2}

\usepackage{CJK}
\usepackage{times}
\usepackage{multirow}
\usepackage[dvipsnames]{xcolor}
\usepackage{amsmath}
\usepackage{amsthm, mathrsfs}
\usepackage{amssymb}
\usepackage{amsbsy}
\usepackage{enumitem}
\usepackage{bm}
\usepackage{wasysym}
\usepackage[english]{babel}
\usepackage[T1]{fontenc}
\usepackage[utf8]{inputenc} 
\usepackage{graphicx}
\usepackage[colorlinks,bookmarks=false,citecolor=blue,linkcolor=red,urlcolor=blue]{hyperref}
\usepackage{pstricks}
\usepackage{rotating}			       
\usepackage{tabularx,hhline}	
\usepackage[caption=false]{subfig}		
\usepackage[normalem]{ulem}

\newcolumntype{P}[1]{>{\centering\arraybackslash}p{#1}}

\begin{document}

\begin{CJK*}{UTF8}{gbsn} 
\title{
Classical $\mathbb{Z}_2$ spin liquid on the 
generalized four-color Kitaev model 
}


\author{Han Yan (闫寒)}
\affiliation{Institute for Solid State Physics, 
The University of Tokyo, Kashiwa, Chiba 277-8581, Japan}
\email{hanyan@issp.u-tokyo.ac.jp}

\author{Rico Pohle}
\affiliation{Institute for Materials Research,
Tohoku University, Sendai, Miyagi 980-8577, Japan}
\email{rico.pohle@tohoku.ac.jp}
\affiliation{Graduate School of Science and Technology, Keio University, 
Hiyoshi, Yokohama 223-8522, Japan}
\affiliation{Department of Applied Physics, The University of Tokyo, Hongo,
Tokyo 113-8656, Japan}
\date{\today}
\begin{abstract}
While $U$(1) spin liquids have been extensively studied in both quantum 
and classical regimes, exact classical $\mathbb{Z}_2$ spin liquids
arising from models with nearest-neighbor, bilinear spin interactions 
are still rare. 
In this Letter, we explore the four-color Kitaev model as a minimal 
model for stabilizing classical $\mathbb{Z}_2$ spin liquids 
across a broad family of tricoordinated lattices. 
By formulating a $\mathbb{Z}_2$ lattice gauge theory, we identify this 
spin liquid as being described by an emergent Gauss's law 
with effective charge-2 condensation, and
deconfined fractionalized bond-charge excitations. 
We complement our findings with Monte Carlo simulations, 
revealing a crossover from a high-temperature paramagnet to a 
low-temperature liquid phase characterized by residual entropy, 
classical $\mathbb{Z}_2$ flux order, and diffuse spin structure factors.
%
\end{abstract}
\maketitle
\end{CJK*}

\textit{Introduction.}
%
Lattice gauge theories, originally introduced by Wegner in 1971~\cite{Wegner1971},
provided the first rigorous mathematical framework for describing
emergent phenomena such as topological order.
Over the years, lattice gauge theories have become essential for understanding
a wide range of exotic phenomena in condensed matter physics,
including the toric code~\cite{Kitaev2003},
$\mathbb{Z}_2$ flux excitations in the Kitaev spin liquid~\cite{Kitaev2006, Baskaran2008},
and resonating valence bond (RVB) liquids~\cite{Moessner2001, Ralko2005, Misguich2002},
bridging diverse topics, from the fundamentals of quantum
computing~\cite{Kitaev2003, Nayak2008} to 
superconductivity~\cite{Senthil2000, Lee2006}.

Spin liquids, strongly correlated systems that defy the conventional Landau 
paradigm of phase transitions,
are typically studied in either their quantum or classical 
regime and are deeply tied to lattice gauge theoretical 
concepts~\cite{Wen2007Book}.
In the classical regime, they exhibit an extensive ground-state 
degeneracy, often described by emergent 
electrostatics~\cite{Davier2023, Yan2024a, Yan2024b, Fang2024PhysRevB}, 
while in the quantum regime, they may host 
phenomena such as topological order, topological entanglement 
entropy, and fractionalized 
excitations~\cite{Balents2010, Savary2017, Zhou2017}.
Although classical spin liquids lack long-range quantum entanglement, 
they can often serve as a foundation for constructing 
quantum spin liquids.
A well-known example is spin ice,
which realizes $U(1)$ Maxwell 
electrostatics~\cite{Harris1997, Moessner1998a, Moessner1998b, SpinIce2021}
and transforms into quantum spin ice upon introducing quantum 
dynamics~\cite{Hermele2004, Shannon2012, Benton2012}.

Building on the fundamental importance of classical spin liquids and 
the recent advances in demonstrating $\mathbb{Z}_2$ quantum spin liquids 
with  Rydberg 
atoms~\cite{Keesling2019Nature, Samajdar2021PNAS, Verresen2021, Semeghini2021, Ebadi2021Nautre}, 
we propose a realistic classical spin model that 
explicitly realizes $\mathbb{Z}_2$ electrostatics.
While many examples of classical $\mathbb{Z}_2$ spin liquids are built 
from dimer liquids that emerge from microscopic spin 
models~\cite{Sachdev1999JPSJ, Moessner2001, Moessner2011, Verresen2022}, 
their direct realization from nearest-neighbor, bilinear spin interactions with an exact 
ground-state degeneracy remains exceedingly rare~\cite{Rehn2017} and 
experimentally challenging to achieve~\cite{Rehn2016}.
Encouragingly, recent semiclassical simulations of the $S=1$ 
Kitaev model, with finite bilinear and biquadratic 
spin interactions, have revealed a novel 
chiral spin liquid~\cite{Pohle2023BBQK_short, Pohle2024BBQK_long}.
This state is characterized by a residual entropy, extremely 
short-ranged spin correlations, and nonzero scalar spin chirality 
marked by $\mathbb{Z}_2$ flux order -- all properties that suggest 
it may represent a classical $\mathbb{Z}_2$ liquid.
However, despite its discovery, a thorough analysis of the underlying 
gauge structure remains an open question.

In this Letter, we present the four-color Kitaev model, which realizes a broad 
family of classical $\mathbb{Z}_2$ spin liquids. 
By mapping spin degrees of freedom to local charges that act as sources of 
electric field fluxes, we show that this spin liquid 
is governed by an emergent Gauss's law with charge mod 2 as the ground state condition.
While different ground states within or across topological sectors are connected through 
loop updates, single-spin flips generate two fractional deconfined flux charges on bonds. 
We argue that the physics generally applies to 
tricoordinated lattices, 
which we explicitly validate on the honeycomb, square-octagon, and star lattices. 
We complement our results by finite-temperature Monte Carlo simulations, which reveal 
an almost lattice-independent crossover from a high-temperature paramagnet to a low-temperature 
liquid phase characterized by residual entropy, classical $\mathbb{Z}_2$ flux order, and diffuse spin 
structure factors.
Our paper provides a detailed explanation of the underlying gauge structure of the 
chiral spin liquid found on the $S=1$ Kitaev model with bilinear-biquadratic interactions, 
as described in Refs.~\cite{Pohle2023BBQK_short, Pohle2024BBQK_long}, and offers 
a framework for studying a broader class of classical $\mathbb{Z}_2$ spin liquids.

%
\begin{figure}[!t]
    \centering
    \includegraphics[width=\columnwidth]{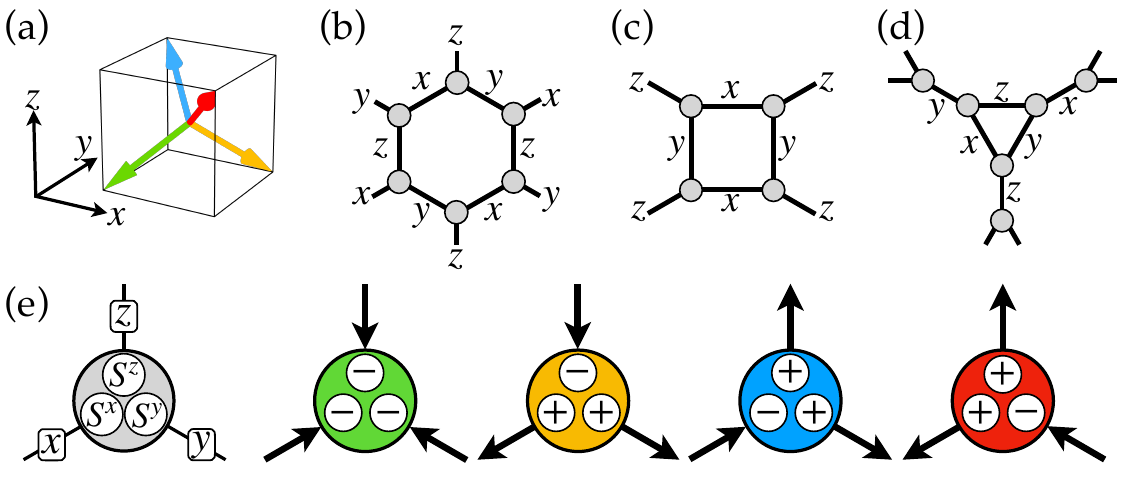}	
    \caption{
    Definition of spin directions in the four-color model with  
    mapping to their effective lattice gauge theory.
    (a) Four considered discrete spin states
    with their corresponding colors.
    Kitaev bond labels are shown for the tricoordinated  
    (b) honeycomb,
    (c) square-octagon, and 
    (d) star lattices. 
    (e) One-to-one mapping of individual spins to their charge 
    degree of freedom. 
    Every site contains three charges with flux lines pointing along 
    the $x$, $y$, and $z$ Kitaev bonds, according to the sign of their 
    respective spin components $S^x$, $S^y$, and $S^z$.
    }
\label{fig:bond.constraint}
\end{figure}

\textit{The four-color Kitaev model.}
%
We consider classical, discretized spins oriented towards four of the 
eight corners of a unit cube, defined as follows:
%
\begin{equation}
	\begin{aligned}
    {\rm green:}  \quad  \boldsymbol \vec{S}_{\rm g} = \tfrac{1}{\sqrt{3}} \left\{ -1, -1, -1 \right \} \, ,   \\
    {\rm yellow:} \quad  \boldsymbol \vec{S}_{\rm y} = \tfrac{1}{\sqrt{3}} \left\{ +1, +1, -1 \right \} \, ,   \\
    {\rm blue:}   \quad   \boldsymbol \vec{S}_{\rm b} = \tfrac{1}{\sqrt{3}} \left\{ -1, +1, +1 \right \} \, ,  \\
    {\rm red:}    \quad   \boldsymbol \vec{S}_{\rm r} = \tfrac{1}{\sqrt{3}} \left\{ +1, -1, +1 \right \} \, .
	\end{aligned}
\label{eq:spin.states}
\end{equation}
%
The  color assignments  (``green'', ``yellow'', ``blue'', and ``red'') and 
spin directions  follow the  convention  
used in Refs.~\cite{Pohle2023BBQK_short, Pohle2024BBQK_long}
and are illustrated in Fig.~\ref{fig:bond.constraint}(a).
The four-color state arises in various contexts: it does not only describe 
the low-temperature physics of the $S=1$ model studied in 
Refs.~\cite{Pohle2023BBQK_short, Pohle2024BBQK_long}, but also serves as 
the simplest discretization of $O(3)$ vector spins. 
Various microscopic models with similar four-dimensional local degree of 
freedom have been  
studied~\cite{Ashkin1943, Kondev1995, Khemani2012, Chern2014, Gomez2024},
making it particularly compelling to investigate its behavior on the Kitaev model.

We consider models with Kitaev-type bond-anisotropic spin interactions 
defined on tricoordinated lattices.
In such lattices every site is connected to its neighbors by 
three distinct types of bonds, typically labeled as $x$, $y$, and $z$, 
ensuring that each site has exactly one bond of each kind.

The Hamiltonian defined on these lattices is
%
%
\begin{equation}
    {\mathscr H}_{\rm 4c} 
		= \sum_{\alpha = x,y,z}\sum_{ \langle ij \rangle_{\alpha}}  
		   S_i^{\alpha} S_j^{\alpha} \ ,
\label{eq:4cModel}
\end{equation}
%
with $S_i^{\alpha}$ being the $\alpha = x, y, z$ component of a discrete vector spin 
at site $i$ in one of the four states shown in 
Eq.~\eqref{eq:spin.states}.
Throughout this paper, we refer to this model as the ``four-color model,'' where 
interactions occur solely for the $\alpha$ spin components 
on the $\langle ij \rangle_{\alpha}$ Kitaev bonds.
We exemplify our study on three tricoordinated lattices in 
two dimensions, 
namely the honeycomb, square-octagon, and star lattices, with 
their corresponding Kitaev-bond labels shown in 
Fig.~\ref{fig:bond.constraint}(b)--\ref{fig:bond.constraint}(d), respectively.

The ground state of the four-color model in Eq.~\eqref{eq:4cModel} is 
obtained by minimizing the energy on each Kitaev bond without  
frustration between different bonds.
Local energies are minimized for multiple configurations 
of color pairs. 
\begin{align}
&\text{On the $x$ bonds:} \nonumber\\
&\qquad  \left( \vec{S}_{\rm g}, \vec{S}_{\rm y} \right) ,  
        \left( \vec{S}_{\rm g}, \vec{S}_{\rm r} \right)  , 
        \left( \vec{S}_{\rm b}, \vec{S}_{\rm y} \right)  , 
        \left( \vec{S}_{\rm b}, \vec{S}_{\rm r} \right)  .
        \label{eq:constraint.x}\\
&\text{On the $y$ bonds:} \nonumber\\
&\qquad \left( \vec{S}_{\rm g}, \vec{S}_{\rm y} \right) ,  
        \left( \vec{S}_{\rm g}, \vec{S}_{\rm b} \right) , 
        \left( \vec{S}_{\rm r}, \vec{S}_{\rm y} \right) , 
        \left( \vec{S}_{\rm r}, \vec{S}_{\rm b} \right) . 
        \label{eq:constraint.y}\\
&\text{On the $z$ bonds:} \nonumber\\
&\qquad \left( \vec{S}_{\rm g}, \vec{S}_{\rm b} \right) ,  
        \left( \vec{S}_{\rm g}, \vec{S}_{\rm r} \right) , 
        \left( \vec{S}_{\rm y}, \vec{S}_{\rm b} \right) , 
        \left( \vec{S}_{\rm y}, \vec{S}_{\rm r} \right) .
        \label{eq:constraint.z} 
\end{align}
%
These configurations set the ground state bond constraints
which are not strong enough 
to enforce long-range order
but instead 
lead to a classical spin liquid state with extensively 
many degenerate ground states, 
as discussed in detail for the eight-color model in 
Ref.~[\onlinecite{Pohle2024BBQK_long}] on the honeycomb lattice.

\textit{Emergent $\mathbb{Z}_2$ gauge structure.}
%
While Eqs.~\eqref{eq:constraint.x}--\eqref{eq:constraint.z} completely 
determine all the ground states of 
${\mathscr H}_{\rm 4c}$ in Eq.~\eqref{eq:4cModel}, the nature of the 
corresponding spin liquids and their field-theoretical interpretation are 
still unknown.  
In the following, we investigate these aspects  
by formulating a $\mathbb{Z}_2$ lattice gauge theory for this spin liquid. 

We construct an equivalent model that treats each spin as a source
of three electric field fluxes.
As illustrated in Fig.~\ref{fig:bond.constraint}(e), we assign fluxes 
to individual Kitaev bonds according to the components
of each spin at the vertex. 
For a spin vector \mbox{$\vec{S} = \{ S^x, S^y, S^z \}$}, 
the assigned flux 
along the $\alpha$ bond is
$\left( \sqrt{3} / 2 \right) S^\alpha$,
whose normalization coefficient $\sqrt{3} / 2$
ensures a net flux change in units of $\pm 1$.
This corresponds to a charge contribution of $\pm 1/2$ to the vertex 
from each spin component, resulting in the total charge 
of $- 3/2$ for $\vec{S}_{\rm g}$ and $+ 1/2$ for 
$\vec{S}_{\rm y}, \vec{S}_{\rm b}$ and $\vec{S}_{\rm r}$.
In visual representations,  we use ``arrows'' to  illustrate  
``electric field fluxes''.

In the ground state, two adjacent sites connected by a bond must also share  
aligned electric field fluxes  
to minimize the bond energy \mbox{$S^\alpha_i S^\alpha_j$}.
This ensures there is no net charge on the bond center, consistent with  
the ground state  constraints  
outlined in Eqs.~\eqref{eq:constraint.x}--\eqref{eq:constraint.z}.
Each ground state is then mapped to a unique configuration of electric 
field fluxes on the lattice, whose vertex charges can be either $-3/2$ or $1/2$.
On the other hand, for every flux 
configuration with zero charge 
on bond centers, there exists a corresponding unique ground-state spin configuration, 
as long as each vertex  conforms to the  
configurations listed in Fig.~\ref{fig:bond.constraint}(e).
While the bond centers remain  
charge-neutral,  each vertex can take a charge of either 
$-3/2$ or $+1/2$.
This condition 
implies the $\mathbb{Z}_2$ Gauss's law at each 
vertex~\cite{Sachdev1999JPSJ},
%
\begin{equation}
    \bm{\nabla} \cdot \bm{E}  =  \text{$-3/2$ or $1/2$} \, .
\label{eq:GaussLaw}
\end{equation}
%
The crucial point is that the allowed charges  
satisfy  the condition  \mbox{$-3/2 = 1/2 \mod 2$}.
This indicates that the system can change its charge by even numbers
and still remains in the ground state, but not by odd numbers. 
Such a concept resembles classical ``charge condensation'', where condensation of 
charge-$2$ particles in $U$(1) gauge theories yields a $\mathbb{Z}_2$ gauge theory, 
which is characterized by gapped 
topological order~\cite{Sachdev1999JPSJ,Kitaev2003,Rehn2017}.  
Consequently, the $U$(1) Gauss's law, $\bm{\nabla} \cdot \bm{E}= 0$, transforms 
to $\bm{\nabla} \cdot \bm{E}= 0 \mod 2$. 
Equation~\eqref{eq:GaussLaw} captures the essence of this law, albeit 
a non-zero background charge of $1/2$, with allowed charges 
being only limited to $-3/2$ and $1/2$, rather than all the numbers 
equivalent to $1/2 \mod 2$.

%
\begin{figure}[!t]
    \centering
    \includegraphics[width=0.7 \columnwidth]{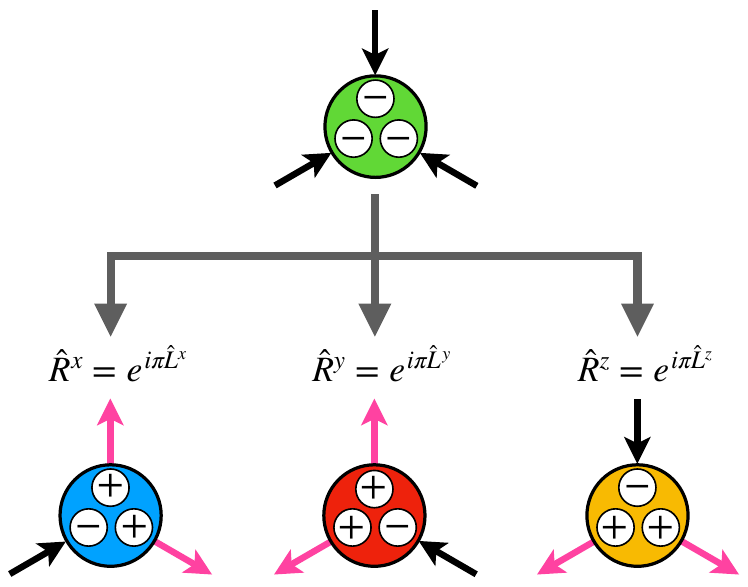}
    \caption{ 
    Single-spin update demonstrated for a green spin $\vec{S}_{\rm g}$
    as defined in Eq.~\eqref{eq:spin.states}. 
    Rotation operators $\hat{R}^{\alpha}$ [see Eq.~\eqref{eq:spin.flip.operator}]
    with \mbox{$\alpha = x,y,z$}, preserve 
    the flux on the $\alpha$ Kitaev bond 
    while inverting the flux on the other two bonds.
    Such an update violates the bond-flux constraint  
    [see Fig.~\ref{fig:bond.constraint}(e) and Eq.~\eqref{eq:GaussLaw}] 
    and creates excitations when acted on a ground state. 
    }
    \label{fig:single.spin}
\end{figure}
%

\textit{$\mathbb{Z}_2$ topological sectors.} 
%
To understand the structure of the topological sectors
in the spin-liquid ground state of ${\mathscr H}_{\rm 4c}$ 
[see Eq.~\eqref{eq:4cModel}] 
we investigate how different ground states
are connected via local and nonlocal loop updates of spins.
We start by discussing single-spin flip updates, as exemplified for $\vec{S}_{\rm g}$  
in Fig.~\ref{fig:single.spin}.
In the spin representation, updating a single spin is equivalent to applying the 
rotation operator 
%
\begin{equation}
    \hat{R}^{\alpha} = e^{i \pi \hat{L}^{\alpha}}  \, ,
\label{eq:spin.flip.operator}
\end{equation}
%
where $\hat{L}^{\alpha}$ is the generator of $SO(3)$ rotations.
The operator $\hat{R}^{\alpha}$ rotates the spin by an angle $\pi$ 
around the $\alpha$ axis,
keeping the spin within the set of 
four color states.

In  the electric field flux picture, $\hat{R}^{\alpha}$ leaves the flux 
on the $\alpha$ bond unchanged while reversing the fluxes 
on the other two bonds.
Note that spins of different color always differ by two of the three electric 
field fluxes. 
Since such a local operation  changes the  
flux on two bonds and creates charges on  their  
bond centers, the new state is always an excited state. 
Therefore, 
single-spin updates do not connect different states in the ground-state manifold.

\begin{figure}[t]
	\centering
        \includegraphics[width= \columnwidth]{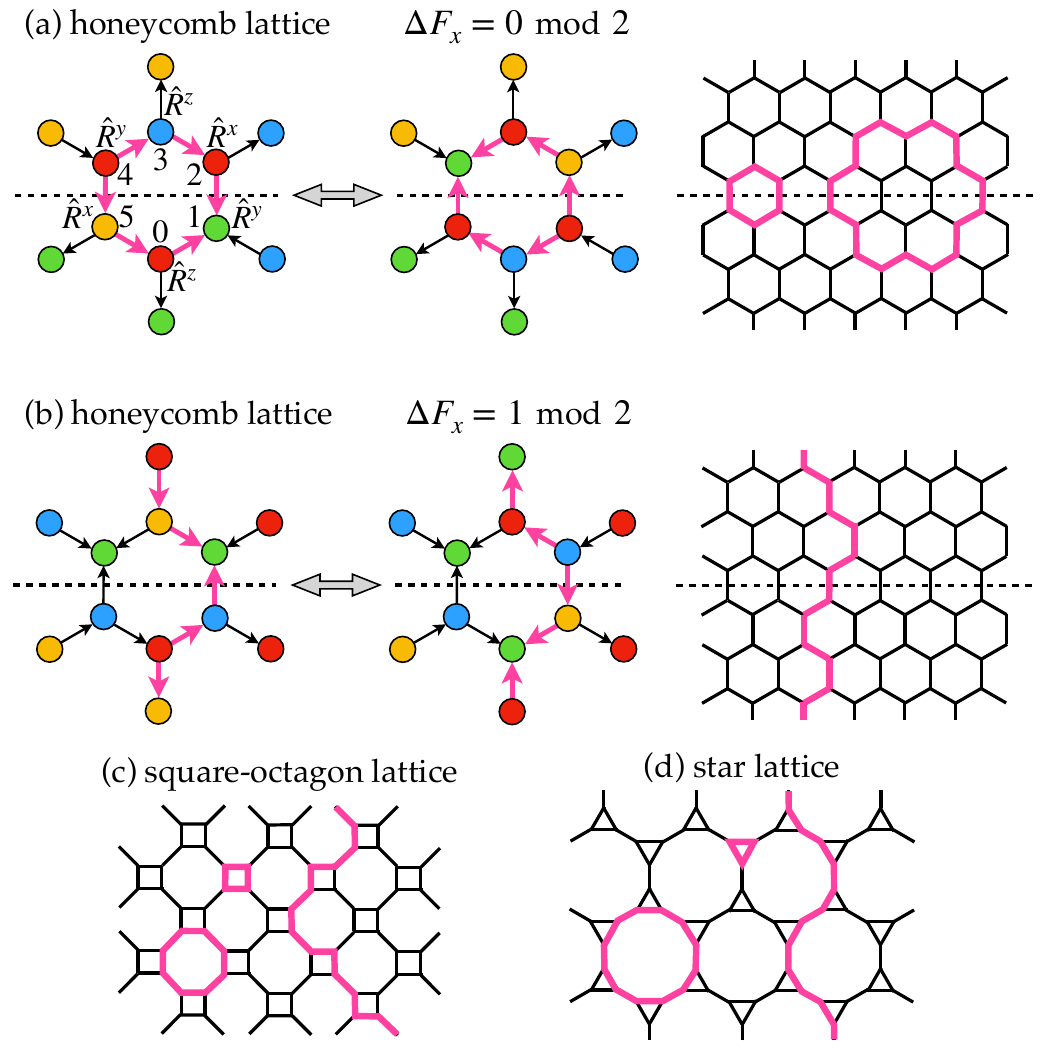}	
	\caption{
        Local and noncontractible loop updates connect states within the 
        spin-liquid ground-state manifold of ${\mathscr H}_{\rm 4c}$ 
        [see Eq.~\eqref{eq:4cModel}].
        These updates involve the successive application of rotation 
        operators $\hat{R}^{\alpha}$ [see Eq.~\eqref{eq:spin.flip.operator}
        and Fig.~\ref{fig:single.spin}] on spins 
        along loops, highlighted in pink. 
        (a) For local loop updates on the honeycomb lattice, 
        the change in flux
        along any closed loop within the bulk 
        (depicted by a dashed-black line)
        takes values of $\Delta F_x =  0  \ \text{mod} \ 2$.
        (b) For noncontractible loop updates, which wrap around the torus of the 
        honeycomb lattice, $\Delta F_x =  1 \ \text{mod} \ 2$, 
        dividing the ground states into different topological sectors.
        Loops are defined in a similar way on other tricoordinated
        lattices, such as (c) the square-octagon lattice
        and (d) the star lattice.
    }
    \label{fig:loop.update}
\end{figure}
To connect different spin configurations within the ground-state manifold, 
one requires multiple  single-spin flips  
in the form of closed loops, 
as shown in explicit examples in Fig.~\ref{fig:loop.update}.
On the honeycomb lattice, the  shortest loop is a hexagon.
Performing such a loop update involves applying 
$\hat{R}^{\alpha}$ [see Eq.~\eqref{eq:spin.flip.operator}] to each of the 
six sites within the hexagon, 
where $\alpha$ is the index of the Kitaev bond pointing outside the hexagon
[see Fig.~\ref{fig:loop.update}(a), left].
The corresponding operator 
%
\begin{equation}
    \hat{W}_p \equiv e^{i\pi \hat{L}_0^z}e^{i\pi \hat{L}_1^y}e^{i\pi \hat{L}_2^x}
                     e^{i\pi \hat{L}_3^z}e^{i\pi \hat{L}_4^y}e^{i\pi \hat{L}_5^x}   \, ,
\label{eq:Wp}
\end{equation}
%
acts locally in the bulk and 
connects different spin-liquid ground states within the same topological sector.
We note that $\hat{W}_p$ in Eq.~\eqref{eq:Wp} is exactly the ``flux operator''  
in the original quantum Kitaev honeycomb model \cite{Kitaev2006, Baskaran2008}, 
if we turn the generators $\hat{L}^\alpha$ into quantum spin 
operators $\hat{S}^\alpha$.
In general, Eq.~\eqref{eq:Wp} can be extended to any tricoordinate 
lattice with
%
\begin{equation}
    \hat{W}_p \equiv \prod_{ \{ j, \alpha \} } e^{i\pi \hat{L}_j^{\alpha}}  \, ,
\label{eq:Wp.general}
\end{equation}
%
where the index $j$ runs over loops of any size,
such as $4$ and $8$ in the square-octagon lattice [see 
Fig.~\ref{fig:loop.update}(c)], or $3$ and $12$ on the 
star lattice [see Fig.~\ref{fig:loop.update}(d)].

To understand the topological sectors of the ground state, we analyze its 
degeneracy structure and explore how flipping of loops can connect different spin 
configurations. 
Given any ground-state configuration, flipping fluxes on a hexagon or 
other local loop always yields another ground state. 
This property arises from allowing the vertex charge to change only by units 
of $\pm2$ [see Eq.~\eqref{eq:GaussLaw}].
Importantly, in these models, local loops are always flippable, and the fluxes do 
not need to connect head-to-tail as is required in $U(1)$ spin 
liquids to maintain a charge of $0$ on every
vertex~\cite{Sachdev1999JPSJ,Huse2003PhysRevLett,Hermele2004}.

We examine the total flux $F_x$ along all bonds intersected by a 
straight line (analogous to a Wilson loop or topological number/logical 
qubit~\cite{Kitaev2003, nielsen00book}) along the $x$ direction on the honeycomb lattice
[dashed line in Figs.~\ref{fig:loop.update}(a) and~\ref{fig:loop.update}(b)].
In Fig.~\ref{fig:loop.update}(a), one observes that the change of flux 
(highlighted in pink) before and 
after the cluster update of a single hexagon is either 
\mbox{$\Delta F_x = \pm 2$} or 
\mbox{$\Delta F_x = 0$} (not shown), 
since the straight line always intersects an even number of bonds. 
Accordingly, the total flux can only change by an even number 
\mbox{$\Delta F_x =  0  \mod 2$} for any 
loop confined within the bulk of the system. 
All ground states connected by such local loop 
updates belong to the same topological sector labeled by 
the net flux \mbox{$F_x \mod 2$}~\cite{Sachdev1999JPSJ,Kitaev2003}.

To alter the total flux $F_x \mod  2$, the loop must
wrap around the torus of the lattice. 
Such a non-contractible loop intersects the straight line an odd number of 
times, as shown in Fig.~\ref{fig:loop.update}(b), and changes the net 
flux by \mbox{$\Delta F_x = 1 \mod  2$}.
Consequently, the system  enters a different topological sector, 
which cannot be reached through local loop updates confined within the bulk. 
The same reasoning applies to other tricoordinated lattices, 
as visualized in Figs.~\ref{fig:loop.update}(c) and \ref{fig:loop.update}(d).
The non-contractible loop-flipping operator is analogous to the 
topological operator/logical gate ~\cite{Kitaev2003,nielsen00book} .

On a torus, the total fluxes $F_{x,y}$ over the cuts in the 
$x$ and $y$ 
directions, being either odd or even, divide the ground states into four 
topological sectors.
States in the same sector are connected via local 
loop updates,
while those in different sectors are not.
This is exactly the classical limit of 
gapped $\mathbb{Z}_2$ topological order~\cite{Kitaev2003}.
In this context, local loop updates act as the analog of 
``magnetic field operators'', 
leaving the system in the ground-state manifold by altering the electric 
field configuration without creating charges on the bond centers.  
On the other hand, noncontractible loop updates take the system 
ground state to a different topological sector~\cite{Dennis2002}.
%

\textit{$\mathbb{Z}_2$ charges.}
%
The lowest-energy excitations  
from the ground state are states with one charge on a bond.
Flipping a single spin inverts the fluxes on two bonds [see Fig.~\ref{fig:single.spin}],
which creates two nonzero charges at the cost of 
$\Delta E = \tfrac{4}{3}.$
Once created, these charges can move independently throughout the system 
via successive single-spin flips without any additional energy cost. 
Thus, such excitations form a classical analog of deconfined 
fractionalized excitations.

These fractional excitations behave like $\mathbb{Z}_2$ charges rather 
than $U(1)$ charges, meaning that only the charge $\mod 2$ is conserved,  
instead of the net charge itself.
One can create two bond-center charges of either $(+1,+1)$, $(+1,-1)$ or $(-1,-1)$ 
by flipping a spin. 
As a charge moves around the lattice through successive single-spin 
flips, its $\pm$ sign can change.
Consequently, for bond-center charges,  only their even/oddness (i.e., charge$\mod 2$) 
is conserved, rather than the total charge.
This characteristic is consistent with the defining feature of a 
$\mathbb{Z}_2$ spin liquid.

\begin{figure}[th]
    \centering
    \includegraphics[width=\columnwidth]{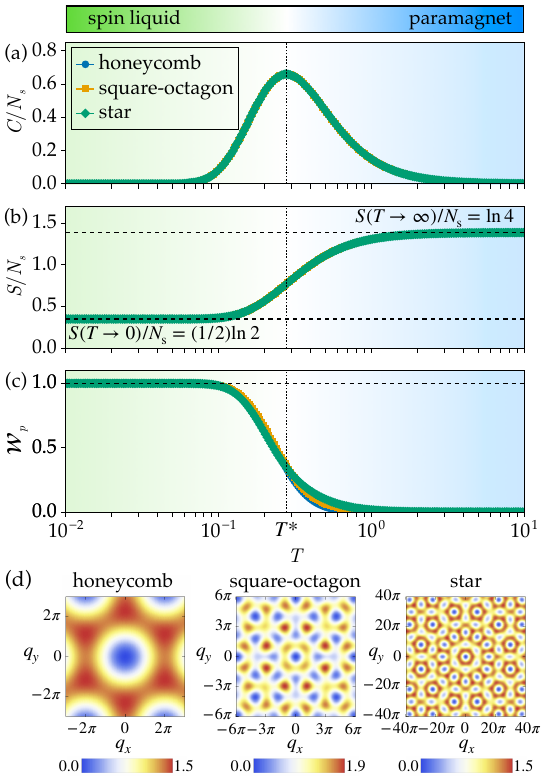} 	
    \caption{	
    Finite-temperature Monte Carlo simulations of ${\mathscr H}_{\rm 4c}$
    [see Eq.~\eqref{eq:4cModel}] on the honeycomb, square-octagon and star lattice,
    reveal a crossover from a 
    high-temperature paramagnet to a low-temperature spin 
    liquid. 
    The panels depict normalized values for
    (a) the specific heat, $C/ N_s$,
    (b) the thermodynamic entropy, $S/N_s$, and
    (c) the classical analog of the $\mathbb{Z}_{2}$-flux operator, 
    ${\mathcal W}_p$ [see Eq.~\eqref{eq:Wp.classical}].
    Panel (d) shows the spin structure factor, $S({\bf q})$, 
    for all three lattice models in the spin-liquid phase
    at $T = 0.01$.
    Thermodynamic observables (spin structure factors) were 
    obtained for finite-size clusters with periodic 
    boundary conditions and linear dimensions of 
    $L = 96$ ($L = 48$) for the honeycomb, 
    $L = 60$ ($L = 24$) for the square-octagon, and 
    $L = 60$ ($L = 12$) for the star lattice models.
    }
\label{fig:thermodyn}
\end{figure}
%

\textit{Finite-temperature properties.}
%
We complement our analytical understanding of the 
spin-liquid ground state in ${\mathscr H}_{\rm 4c}$ 
[see Eq.~\eqref{eq:4cModel}] with finite-temperature 
Monte Carlo (MC) simulations. 
As discussed previously, single-spin flip updates alone are insufficient  
to thermalise the system at low temperatures.
Therefore, we use a hybrid Monte Carlo scheme that
combines single-spin flip updates with local loop updates. 
In this scheme, a single MC step consists of $N_s$ (total site number)
local Metropolis spin-flip attempts at randomly chosen sites of the lattice,  
followed by $N_l$ (total number of elementary loops) cluster update 
attempts by applying Eq.~\eqref{eq:Wp.general} 
on randomly chosen elementary loops on the lattice.
To further mitigate correlation times, we employ parallelization 
in temperature using the replica-exchange algorithm \cite{Swendsen1986} every 
100 MC steps.
Thermodynamic quantities are averaged over $5 \times 10^5$ statistically 
independent samples, after $1 \times 10^6$  steps of simulated annealing and 
thermalization each.

In Fig.~\ref{fig:thermodyn}, we show finite-temperature MC simulation 
results of ${\mathscr H}_{\rm 4c}$ [Eq.~\eqref{eq:4cModel}] 
for honeycomb, square-octagon and star lattice models.
The specific heat 
exhibits a lattice independent crossover from a high-temperature 
paramagnet to the low-temperature spin liquid with 
a broad peak spanning over one order of magnitude in temperature,
and a peak maximum at $T^* = 0.272(8)$.
We note that the four-color model itself explicitly breaks the 
time-reversal symmetry by considering only four allowed spin states, 
as illustrated in Fig.~\ref{fig:bond.constraint}(a).
Consequently, we do not observe a finite-temperature phase 
transition, in contrast to 
the observations in Ref.~\cite{Pohle2024BBQK_long}, which investigates 
the time-reversal symmetric case involving eight states. 

Upon cooling, the system releases its thermodynamic entropy, 
\mbox{$S(T \to \infty) / N_s = \ln{4}$},
to a residual value of 
\mbox{$S(T \to 0) / N_s = \tfrac{1}{2} \ln{2}$}. 
This residual entropy reflects the remaining degeneracy in the ground state, 
counting always two possible configurations for every elementary loop in 
the lattice. 
A similar behavior is observed in the classical analog of the
$\mathbb{Z}_{2}$-flux operator ${\mathcal W}_p$,
%
\begin{equation}
    {\mathcal W}_p = \prod_{j \in p} \sqrt{3} S^{\alpha}_j   \, , 
\label{eq:Wp.classical}
\end{equation}
%
which is not an operator per se but rather a quantity that evaluates  
whether the Gauss's law [Eq.~\eqref{eq:GaussLaw}] is globally 
satisfied or not.
Upon cooling, ${\mathcal W}_p$ gradually increases within the crossover window, 
eventually reaching a value of $+1$  at 
low temperatures. 

Characteristic magnetic scattering signatures 
in the spin-liquid phase
of all three lattice models
are shown in the energy-integrated structure factors
$S({\bf q})$ in Fig.~\ref{fig:thermodyn}(d).
The signal for all lattice models is very diffuse, 
directly indicating the absence of any conventional
magnetic order.
The absence of singularities, such as ``pinch-points'', characteristic 
of Coulombic $U(1)$ liquids, stems from the 
underlying $\mathbb{Z}_2$ Gauss law in Eq.~\eqref{eq:GaussLaw}.
Consequently, this spin liquid belongs to the category of fragile topological spin 
liquids~\cite{Yan2024a, Yan2024b}, characterized by a gapped spectrum 
of the interaction matrix and 
exponentially decaying spin-spin correlations. 
Notably, these spin-spin correlations are extremely short-ranged, 
aligning with analytical 
predictions for the $S=1$ Kitaev spin liquid in Ref.~\cite{Baskaran2008}.
While the honeycomb lattice exhibits a periodic structure with dominant weight 
concentrated near the Brillouin zone edges,  
the square-octagon and star lattices display an
aperiodic pattern. 
This aperiodicity originates from the fractional distances of real-space 
lattice sites
and has also been observed in other spin liquids, e.g., on the 
ruby lattice~\cite{Rehn2017} or the square-kagome lattice~\cite{Pohle2016}.
%

{\it Conclusions and discussion.}
%
We have studied the generalized four-color Kitaev model 
and demonstrated that it stabilizes a classical $\mathbb{Z}_2$ 
spin liquid across a broad family of tricoordinated lattices.
By developing an effective $\mathbb{Z}_2$ lattice gauge theory for 
this family of models, we identified an emergent Gauss's law, constrained to 
charge mod 2, as the ground state condition, leading to the formation of 
topological sectors and deconfined bond-charge excitations.

Our paper offers a significant contribution to the search for exotic spin liquids, 
particularly those exhibiting Kitaev-type anisotropies. 
Moreover, it represents a rare example of a $\mathbb{Z}_2$ classical liquid 
directly realized in a model with bilinear spin interactions, rather than as 
an emergent form of a dimer liquid or from models with multi-spin interactions.

The most promising place to look for a realization of this 
$\mathbb{Z}_2$ spin liquid may lie in $S=1$ Kitaev models on the 
honeycomb lattice with additional bilinear-biquadratic spin interactions, 
as studied in Refs.~\cite{Pohle2023BBQK_short, Pohle2024BBQK_long}.
In those studies, semiclassical simulations demonstrated that
the right balance of dominant Kitaev and finite bilinear-biquadratic 
interactions stabilizes a  finite-temperature chiral spin liquid.
The physics of this liquid can be effectively captured by the four-color model 
given in Eq.~\eqref{eq:4cModel}.
Potential experimental realizations might be found in honeycomb 
materials composed of Ni$^{2+}$ ions~\cite{Stavropoulos2019}, 
where positive biquadratic 
interactions may arise from orbital 
degeneracy~\cite{Yoshimori1981, Hoffmann2020, Soni2022}.

While the conclusions presented in this paper generally apply to 
tricoordinated 
lattices, we have made our discussions explicit on only three lattice models 
in two dimensions: 
the honeycomb, square-octagon, and star lattice. 
We believe an extension  
to tricoordinate lattices in three dimensions is 
straightforward and offers another promising route to discover exotic spin
liquids~\cite{Mandal2009, OBrien2016, Eschmann2020, Mishchenko2020}.

An important direction
for future investigations is to explore the 
connection between our results and quantum models at zero temperature.
The results in this paper, along with those in
Refs.~\cite{Pohle2023BBQK_short, Pohle2024BBQK_long}, primarily  
address classical physics, 
where quantum entanglement between spins is absent.
While  local perturbations in such classical models lift 
the degeneracy and lead the system into a long-range ordered 
phase, 
incorporating quantum dynamics---for example, through
local loop operators such as the $\mathbb{Z}_{2}$-flux operator
${\mathcal W}_p$ in Eq.~\eqref{eq:Wp.general}---could 
potentially drive the system into topological 
order~\cite{Verresen2022, Tarabunga2022},
which would be robust against perturbations. 
Furthermore, the connection between our results and 
quantum $S=1$ Kitaev models~\cite{Baskaran2008, Hickey2020, Zhu2020, Dong2020, Lee2020, Khait2021, Fukui2022, Georgiou2024, Ralko2024, Ma2023, Liu2024arXiv}, 
chiral spin liquids on tricoordinated 
lattices~\cite{Yao2007, Dusuel2008, Nasu2015, Ralko2020, Khait2021, Hickey2021, Ornellas2024},
quantum loop models~\cite{Savary2021}, and 
$\mathbb{Z}_4$ Kitaev spin liquids~\cite{yang2024chiralspinliquidgeneralized},
remains 
an open and fascinating question for future studies.
In particular, we hope our results will
provide valuable insights into
the concrete construction 
of ground-state wave functions for 
high-spin quantum Kitaev models and their 
variants. 
%

\begin{acknowledgments}
The authors are indebted to Yukitoshi Motome and Nic Shannon for helpful 
conversations and a critical reading of the manuscript.
R.P. acknowledges support from the 
Quantum Liquid Crystals JSPS KAKENHI Grant No. JP19H05825, and
MEXT as ``Program for Promoting Researches on the Supercomputer Fugaku'' 
(Grant No. JPMXP1020230411).
H.Y. is supported by the 2024 Toyota Riken Scholar Program from the Toyota Physical 
and Chemical Research Institute, and the  Grant-in-Aid for Research Activity Start-up from Japan Society
for the Promotion of Science (Grant No. 24K22856).
Numerical calculations were carried out using HPC facilities provided 
by the Supercomputer Center of the Institute for Solid State Physics, 
the University of Tokyo.
\end{acknowledgments}

\bibliography{Bibliography}

\end{document}